\documentstyle[12pt,epsf]{article}
\def\beq{\begin{equation}}
\def\eeq{\end{equation}}
\def\beqa{\begin{eqnarray}}
\def\eeqa{\end{eqnarray}}
\def\d{{\rm d}}

\begin{document}

\begin{flushright}
ITP--SB--96--54
\end{flushright}


\centerline{\Large \bf PQCD Analysis of Hard Scattering in Nuclei\footnote{Talk
presented at the RHIC Summer Study '96, Brookhaven National Laboratory, July 8-18, 1996.}}
\vspace*{.5in}

\centerline{Jianwei Qiu$^a$ and George Sterman$^b$}
\vspace*{.25in}

\centerline{\small \it $^a$ Department of Physics and Astronomy, 
Iowa State University}

\centerline{\small \it Ames, Iowa 50011}

\centerline{\small \it $^b$Institute for 
Theoretical Physics, State University of
New York}

\centerline {\small \it Stony Brook, NY 11794-3840}
\vspace*{.5in}

\begin{abstract}
We review the extension of the factorization formalism
for perturbative QCD to soft initial- and final-state
scattering associated with hard processes in nuclei.
\end{abstract}

\section{Introduction}

\begin{figure}[ht]
\centerline{\epsffile{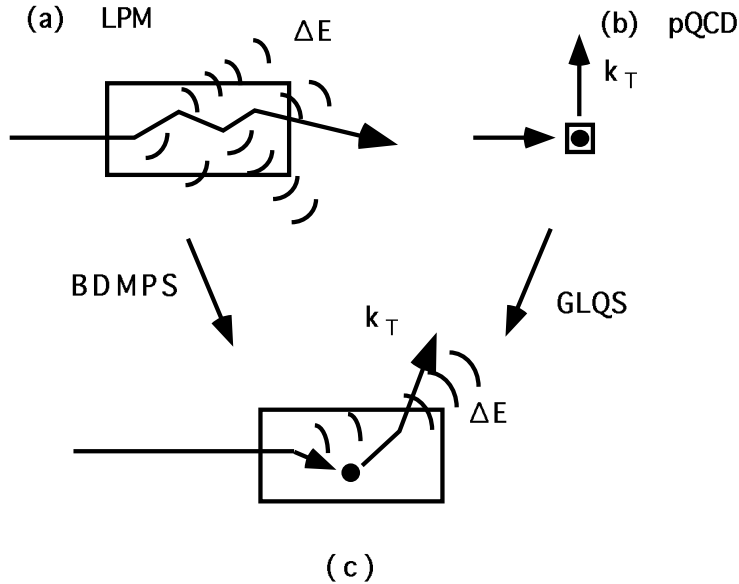}}
\caption{Alternate approaches to hard scattering in nuclei. (a) Landau-Pomeranchuk-Migdal 
analysis treats energy loss due to many soft scatterings. (b) Perturbative QCD analysis
treats momentum transfer due to hard scattering. (c)  For scattering in nuclei,
both must be combined.}
\label{fig1}
\end{figure}

In this talk, we would like to review a few results from a perturbative QCD (pQCD)
treatment of the scattering of hadrons and leptons in nuclei, based on factorization,
work in collaboration with Ma Luo \cite{LQS1,LQS,LQS3} and, more recently,
Xiaofeng Guo \cite{GQ}.
At the outset, it may be useful to clarify the relation of this work
to the recent papers of Baier {\it et al.} (BDMPS), described by 
Dokshitzer at this workshop \cite{BDMPS}.  We have tried to illustrate this
relation schematically in Fig.\ \ref{fig1}.  
The BDMPS analysis begins (Fig. 1a) with the classic treatment of 
radiation
induced when a charged particle passes through a large target, 
due originally
to Landau, Pomeranchuk and Migdal (LPM).  This analysis does not require
the presence of  a hard scattering, but describes
the coherent results of  many soft scatterings.  Its primary subject 
has traditionally been induced energy loss.  Our analysis (GLQS) begins with the
perturbative QCD treatment of hard-scattering in a small target (Fig.\ 1b),
in which the primary subject of interest is momentum transfer.  
A complete analysis (Fig.\ 1c) of hard scattering
in a large target, involves both energy loss and the transverse momenta due to 
initial- and final-state soft scatterings.  Our work is a  step in
this direction, attempting to stay as close as possible to 
the pQCD formalism, in which we may readily quantify corrections.  
To be specific, we consider only a single soft inital- or final-state interaction
in addition to the hard scattering.  
Our central observation is that 
for suitably-defined jet  and related inclusive
cross sections this is the first order in an
expansion in the quantity
\beq
{A^{1/3}\times\lambda^2\over Q^2}\, ,
\label{param}
\eeq
where $\lambda$ represents a nonperturbative scale,
which we shall identify  with a higher-twist 
parton distribution below.  
That additional scatterings are suppressed by factors of  $1/Q^2$ is perhaps
surprising.  Let us review why this is the case, at least for certain 
cross sections. 

The basic analysis of hard-scattering in nuclear matter  (cold or hot) 
\cite{tdilat} is quite simple.  To be specific, consider the
scattering of a quark.  A hard-scattering with momentum transfer
$Q$ can resolve states whose lifetimes are as short as $1/Q$, for instance
quarks off-shell by order $Q$, but still less that $Q$. 
The off-shellness of the scattered quark increases with the momentum
transfer simply because {\it the number of available states
increases with increasing momentum}.
Similarly, the scattered  quark, of momentum $p'$ is typically
off-shell by order $m_J\le Q$.  
We may think of $m_J$ as  the momentum of
the jet into which quark fragments.  
If we are to recognize the jet, we must have $m_J\ll E_J=p'_0$.  On the other
hand, the counting of available states ensures that $m_J\gg\Lambda_{\rm QCD}$.  
 
Now the scattered quark
has a lifetime in its own rest frame $\Delta t^{(p')} \sim {1\over m_J}$
with $m_J\ll E_J$.
In the target rest frame, however, this becomes, for large enough $E_J/m_J$,
$\Delta t^{\rm (target)} \sim{1\over m_J}\left({E_J\over m_J}\right)>R_A$,
where $R_A$ is the (fixed) target size.  Thus, at high enough energy
the lifetime of the scattered quark will exceed the target size, even
though the quark itself is far off the mass shell, typically 
by a scale that grows with the momentum transfer $Q$.

Further couplings of the off-shell quark are suppressed, first of all by the
strong coupling evaluated at scale $m_J$, and, more importantly,
by an overall factor of $1/m_J^2\sim 1/Q^2$, since the effective
size of the scattered quark decreases with momentum
transfer in this manner.  

In summary, for inclusive processes such as jet production, high-$Q$ implies that
process-dependent multiple
scattering is power-suppressed compared 
to single scattering.  Initial-state interactions
internal to the nucleus are leading-power, but factorize.
Thus the ``Cronin effect", $A^\alpha$-dependence
with $\alpha>1$, due to 
multiple scattering, is higher-twist for inclusive
distributions, while the ``EMC" effect for parton
distributions in nuclei is (almost by definition)  
leading-twist.

The most important point here is that the scattered particle
remains off-shell for its entire transit of the target.
Thus, its interactions with the target may be 
treated by the formalism of  perturbative QCD,
which, however, must be extended to include corrections
that decrease with extra powers of momentum transfer.
Up to the first such ``higher-twist" contribution, a general
cross section has the
representation \cite{QS}
\begin{equation}
\sigma(Q) = H^0\otimes f_2\otimes f_2 +
\left(\frac{1}{Q^2}\right) H^1\otimes f_2 \otimes f_{4}
+ O\left(\frac{1}{Q^4}\right)\, ,
\label{HTE}
\end{equation}
where $\otimes$ represents covolutions in fractional momenta
carried by partons, and $f_n$ represents a parton distribution of
twist $n$.
Target-size dependence due to multiple scattering can
only appear in the second term in this expansion.

\section{Parton-Nucleus Scattering in Perturbative QCD}

\subsection{Factorization at Leading and Nonleading Powers}

Let us review some of the details of a factorized cross section like (\ref{HTE}).
The first term, consisting of only twist-two matrix elements has the detailed
form,
\beq
\omega{d\sigma_2\over d^3p'}
=
\sum_{ij}\int{dy}f_{j/p_2}(y,Q)\; \int{dx}f_{i/p_1}(x,Q)
\, \hat{\sigma}_{ij}(xp_1,yp_2,p')\, ,
\label{twist2conv}
\eeq
where we may take $p'$ as the momentum of  an observed jet.
The fragmentation of a jet, suitably defined, is calculable in
perturbation theory, and may be absorbed into the ``hard scattering function"
$\hat{\sigma}$.  The $f_{a/p}$ are distributions of parton type $a$  in
hadron $p$.  They have the interpretaion of expectation
values in the hadronic state
of products of fields on the light cone, for
instance, for a quark distribution
\beq
f_{q/p}(x,Q)=
\int{dy^- \over 2\pi} {\rm e}^{ixp^+y^-}
\langle p|\bar{q}(0){\gamma^+\over 2}q(y^-)|p\rangle\, ,
\eeq
where for simplicity we choose the $A^+=0$ gauge, assuming $\vec{p}$ is
in the plus direction.  Eq.\ (\ref{twist2conv}) is ilustrated
by Fig.\ 2a.   
\begin{figure}[ht]
\centerline{\epsffile{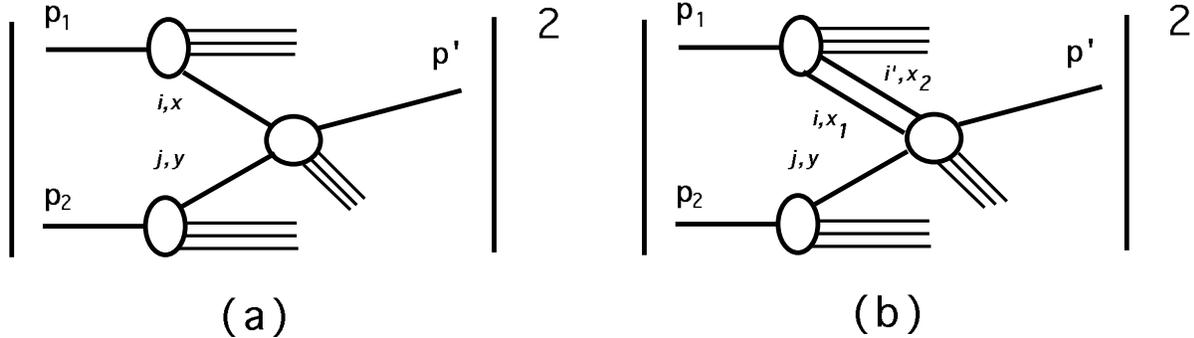}}
\caption{Perturbative QCD at leading twist (a), and higher twist (b).}
\label{fig2}
\end{figure}
As shown, the convolution in eq.\ (\ref{twist2conv}) is in terms of the
momentum fractions $x$ and $y$ carried by partons $i$ and  $j$,
from hadrons $p_1$ and $p_2$, respectively, into the hard scattering.

Fig.\ 2b is the corresponding picture for a higher-twist contribution
to hard scattering.  In this case {\sl two} partons $i$ and $i'$
with momenta $x_1p_1$ and $x_2p_1$ from the
target (the ``nucleus")  collide with a single parton $j$ of
momentum $yp_2$ (from the ``projectile"), 
\beq
\omega{d\sigma_4\over d^3p'}
=
\sum_{(ii')j}\int{dy}f_{j/p_2}(y,Q) 
\, \int dx_1 dx_2 dx_3\; T_{(ii')/p_1}(x_1,x_2,x_3,Q)\,
\hat{\sigma}^{(4)}_{(ii')+j}(x_ip_1,yp_2,p')\, .
\label{sig4}
\eeq
 The expectation value $T$ 
corresponding to this multiparton contribution from the 
target is typically of the form \cite{QS},
\beq
T_{(ii')/p}(x_1,x_2,x_3,Q)= \int{dy^-_1dy^-_2dy^-_3\over(2\pi)^3}
{\rm e}^{ip^+(x_1y^-_1+x_2y^-_2+x_3y^-_3)}
\langle p|B^\dagger_{i}(0)B^\dagger_{i'}(y^-_3)B_{i'}(y^-_2)B_{i}(y^-_1)
|p\rangle\, ,
\label{Tdef}
\eeq
where $B_i$ is the field corresponding to a parton of type $i=q,{\bar q},G$.
In eq.\ (\ref{sig4}), 
the hard part $\hat{\sigma}^{(4)}_{(ii')+j}$  depends on the
identities and momentum fractions of the incoming partons,
but is otherwise independent of the structure -- in particular
the size -- of the target (and projectile).  To find $A$-enhancement due
to multiple scattering, we must look elsewhere. 

\subsection{A-Enhancement from Matrix Elements}

\begin{figure}[ht]
\centerline{\epsffile{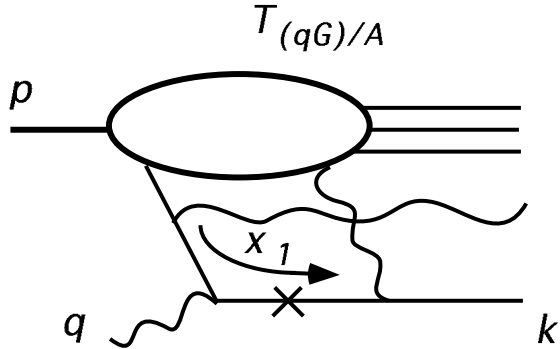}}
\caption{Pole that gives rise to an $A$-enhanced cross section.}
\label{fig3}
\end{figure}

For definiteness, we  consider photoproduction or deeply inelastic scattering on
a nucleus \cite{LQS1,LQS3}.  In this case, the additional soft scattering is
always a final-state interaction.
The structure of the target is manifest only in 
the matrix element $T$ in eq.\ (\ref{sig4}). 
Each pair of fields in the matrix element (\ref{Tdef}) represents
a parton that participates in the hard scattering.
The $y^-_i$ integrals parameterize the distance between
the positions of these particles along the path of the
outgoing scattered quark.  
In eq.\ (\ref{Tdef}), integrals over the
distances $y^-_i$ generally cannot grow with the size
of the target because of oscillations of the exponential factors
${\rm e}^{ip^+x_iy^-_i}$.  Poles from $\hat{\sigma}$ in the $x_i$ integrals, 
associated
with the scattered particle, however, can result in 
finite contributions from points where two of the $x_i$ vanish [1-3].  
An example is shown in Fig.\ \ref{fig3}.
It is important to emphasize that using
a pole in the complex $x_i$ (longitudinal momentum) 
space to do the integral does not correspond to 
assuming on-shell propagation for the 
scattered quark.  Indeed, the $x_i$ integrals are not pinched
between coalescing singularities at that  point, and the same
results could be derived by performing the $x_i$ integrals
without ever going through the $x_i=0$ points.

The result of this reasoning is that matrix elements that depend
on three fractional momenta, as in (\ref{Tdef}) above, 
simplify to a form like
\begin{eqnarray}
T_q(x,A) &=& \int {\d y^-_1 \over 2\pi} 
e^{ip^+xy_1^-}
\int {\d y^- \d y^-_2 \over 2\pi}
\theta(y^--y_1^-)\theta(y_2^-) \nonumber
\\  
&\ &\quad \times\; {1 \over 2}\; \langle p_A|{\bar q}(0)
\gamma^+ F^{\alpha +}(y_2^-)
{F^{+}}_\alpha(y^-) q(y_1^-)|p_A\rangle\ , 
\label{qkme}
\end{eqnarray}
where $|p_A\rangle$ is the relevant nuclear state.  In this
form, integrals over the $y_i^-$ can grow with the nuclear
radius as fast as $A^{1/3}$, once local color confinement is taken
into account.  The variable $x$ here is the
fractional momentum associated with the hard parton
from the target that initiates the process.
The soft scattering contributes a negligible longitudinal fractional momentum.
Details of the reasoning and calculation for deeply inelastic scattering
are given in Ref.\ \cite{LQS3}.

\section{Applications}

In Refs.\ \cite{LQS1} and
\cite{LQS3}, we have applied the formalism sketched above to single-particle
inclusive and single-jet production for
deeply inelastic scattering and photoproduction.  These
cases involve final-state interactions only. 
In each case, the leading $1/Q^2$ correction
is proportional to the matrix element  in eq.\ (\ref{qkme}), or
to a corresponding matrix element $T_G$ with four gluon fields.  
Of course, the value of the correction cannot be estimated
without an idea of the magnitudes of the $T$'s.  Since these magnitudes
are nonperturbative they must be taken from experiment.
At the same time, we expect the $x$-dependence of
the probability to detect  the hard parton to
be essentially unaffected by the presence or  absence of an
additional soft scattering.  Thus, we choose ansatz 
\beq
T_q(x,Q)=\lambda^2 A^{1/3}f_{q/p_A}(x,Q)
\label{Tqans}
\eeq
for $T_q$, in terms of the corresponding twist-two parton distribution $f$,
with $\lambda$ a constant with dimensions of mass (see eq.\ ({\ref{param})).
This assumption facilitates the comparison to data.

A quantity that is sensitive to final-state rescattering in a particularly
direct way is the momentum imbalance of di-jets in  photoproduction in
nuclei.  The $A^{4/3}$ dependence of this 
 quantity is related to the matrix elements $T_q$ and
$T_G$ by the simple formula \cite{LQS}
\beq
\langle k_T^2 E_\ell {\d \sigma \over \d^3\ell} \rangle_{4/3} 
=
\sum_{a=q,g} \int \d x\, T_a(x,A)\, H^{\gamma a}(xp,p_\gamma,\ell)
=
\lambda^2 
A^{4/3} \sum_{a=q,g} \int \d x f_a(x,A) H^{\gamma a}(xp,p_\gamma,\ell)\, ,
\label{dsigma}
\eeq
where $H^{\gamma a}$ is a hard-scattering function that we have computed to lowest
order and where in the second equality we have used (\ref{Tqans}).  The
momentum $\ell$ may be identified as the momentum of the more energetic
jet.
By comparing eq.\ (\ref{dsigma}) to data, \cite{E683} we 
found $\lambda^2\sim 0.05-0.1$ GeV$^2$ \cite{LQS}.  This value
may be used to predict anomalous $A$-enhancement for other
processes.

One such process is direct photon production at measured  transverse
momentum, whose very moderate $A$-dependence has been measured by the
E706 experiment at Fermilab.  In Ref.\ \cite{GQ}, it was
found that the value of $\lambda^2$ above, which produces a relatively
large enhancement in 
dijet momentum imbalance, due to final-state interactions,
produces a quite small $A$-enhancement 
in photoproduction, due to initial-state interactions,
consistent with experiment.  This may
shed some light on the long-standing observation that (initial-state) transverse
momentum effects in Drell-Yan cross sections are also surprisingly
small \cite{DY}.  Clearly, further study of this and related questions is
in order.


\begin{thebibliography}{99}

\bibitem{LQS1} M.\ Luo, J.\ Qiu and G.\ Sterman, {Phys. Lett.}\
{B279} (1992) 377.

\bibitem{LQS} M.\ Luo, J.\ Qiu and G.\ Sterman, Phys.\ Rev.\ D49 (1994) 4493.

\bibitem{LQS3} M.\ Luo, J.\ Qiu and G.\ Sterman, Phys.\ Rev.\ D50 (1994) 1951.

\bibitem{GQ} X.\ Guo and J.\ Qiu, Phys.\ Rev.\ D53 (1996) 6144. 

\bibitem{BDMPS} R.\ Baier, Yu.L.\ Dokshitzer, A.H.\ Mueller, S.\ Peign\'e
and D.\ Schiff, hep-ph/9607355 and 9608322.

\bibitem{tdilat} A.H.\ Mueller, in {\it Proceedings of the XVII
Rencontre de Moriond, Vol.\ 1}, ed.\ J.\ Tran Thanh Van,
(Editions Frontieres, Gif-sur-Yvette, France, 1982).

\bibitem{QS} J.\ Qiu and G.\ Sterman, Nuc.\ Phys.\ B353 (1991) 105, 137.

\bibitem{E683}  D.\ Naples {\it et al.}\ (E683 Collaboration)
Phys.\ Rev.\ Lett.\ 72 (1994) 2341.

\bibitem{DY} M.L.\ Swartz {\it et al.}\ Phys.\ Rev.\ Lett.\ 53 (1984) 32;
D.M.\ Alde {\it et al.\ ibid} 66 (1991) 2285.

\end{thebibliography}
\end{document}